\documentstyle[11pt,newpasp,twoside,epsf]{article}

\def\edcomment#1{\iffalse\marginpar{\raggedright\sl#1\/}\else\relax\fi}
\reversemarginpar

\begin{document}

\title{Chandra Observations of Nearby Spiral Galaxies}
\author{Andrea H. Prestwich}
\affil{Harvard-Smithsonian Center for Astrophysics, 60 Garden St. MS4,
Cambridge, MA 02138}

\begin{abstract}
Chandra is revolutionising our understanding of the X-ray emission
from spiral galaxies.   Before Chandra, detailed extragalactic studies
were limited to Local Group galaxies.  We are now able to obtain
detailed inventories of sources in galaxies outside the Local Group,
allowing  population studies for the first time.  In this
review I will first discuss some intriguing observations of globular
cluster sources in M31, then move outside the Local Group and
summarise the observational properties of sources in several spiral
galaxies (both bulge dominated and disk dominated).  The X-ray colors
and variability characteristics suggest that most of these sources are
accreting binaries. High mass X-ray binaries are formed preferentially
in disk galaxies.  Finally, I show that the luminosity functions of
spiral galaxies can be qualitatively understood in terms of simple
population models.
\end{abstract}

\section{Introduction}
This article focuses on how Chandra's superb angular resolution has
improved our understanding of the X-ray emission from spiral
galaxies, especially the discrete source population.  I would
therefore like to start by briefly reviewing the pre-Chandra view of spiral
galaxies, and explain why Chandra is so crucial for population studies. 
The pre-Chandra view of Local Group galaxies was quite detailed.  This
is demonstrated in Figure~1, which  shows
the PSPC image of the 
Local Group spiral M31 (see Supper et al 2001).  A total of 560
sources are detected in a 10.7 deg$^2$ field of view down to a
limiting sensitivity of 10$^{35}$-10$^{36}$ ergs s$^{-1}$.  The bulge
region is very confused, and contains many unresolved sources and
diffuse emission.  There are also bright sources associated with the
spiral arms.  Many of the detected X-ray sources have optical
counterparts, including 55 stars, 33 globular clusters, 16 supernova
remnants and 10 background sources.     

\begin{figure}
\caption{PSPC image of Local Group galaxy M31 from Supper et al 2001.}
\plotfiddle{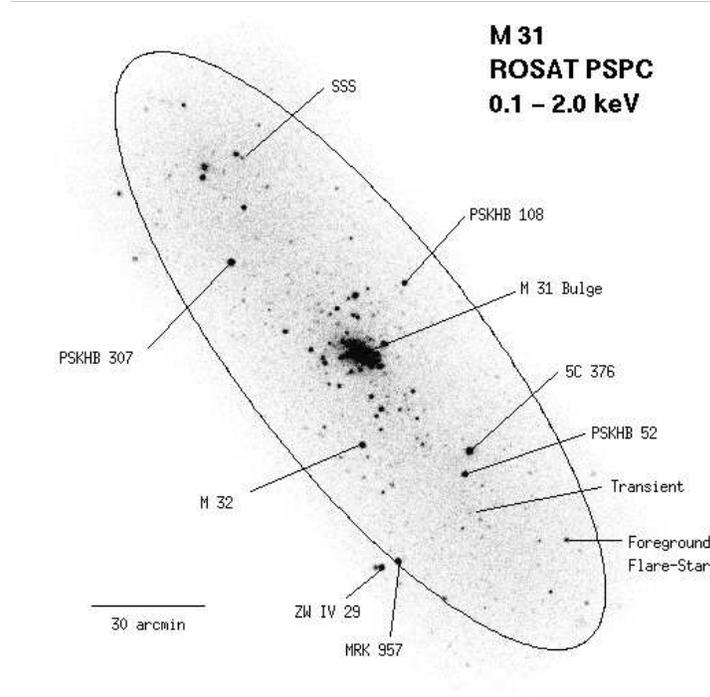}{2.75in}{0.0}{60}{60}{-180.0}{-230.0}
\vskip 0.75in
\end{figure}

Outside of the Local Group, however, the resolution of pre-Chandra
telescopes limited detection to all but the brightest discrete
sources, and confusion was always a problem   The poor resolution also
results in large positional uncertainties, making optical
identifications difficult (see Section~\ref{hyper}).   The improvement
in resolution with the advent of Chandra is illustrated by comparing
the ROSAT HRI observation of M101 (Wang et al 1999) with the recent
Chandra paper on the same galaxy by Pence et al 2001 (e.g. see the
Figure~2.)  Wang et al
(1999) detected a total of 51 sources in the HRI field of view (30
arcmin), whereas Pence et al 2001 detected 110 sources on the S3 chip
alone (field of view 8 arcmin).

\begin{figure}
\plottwo{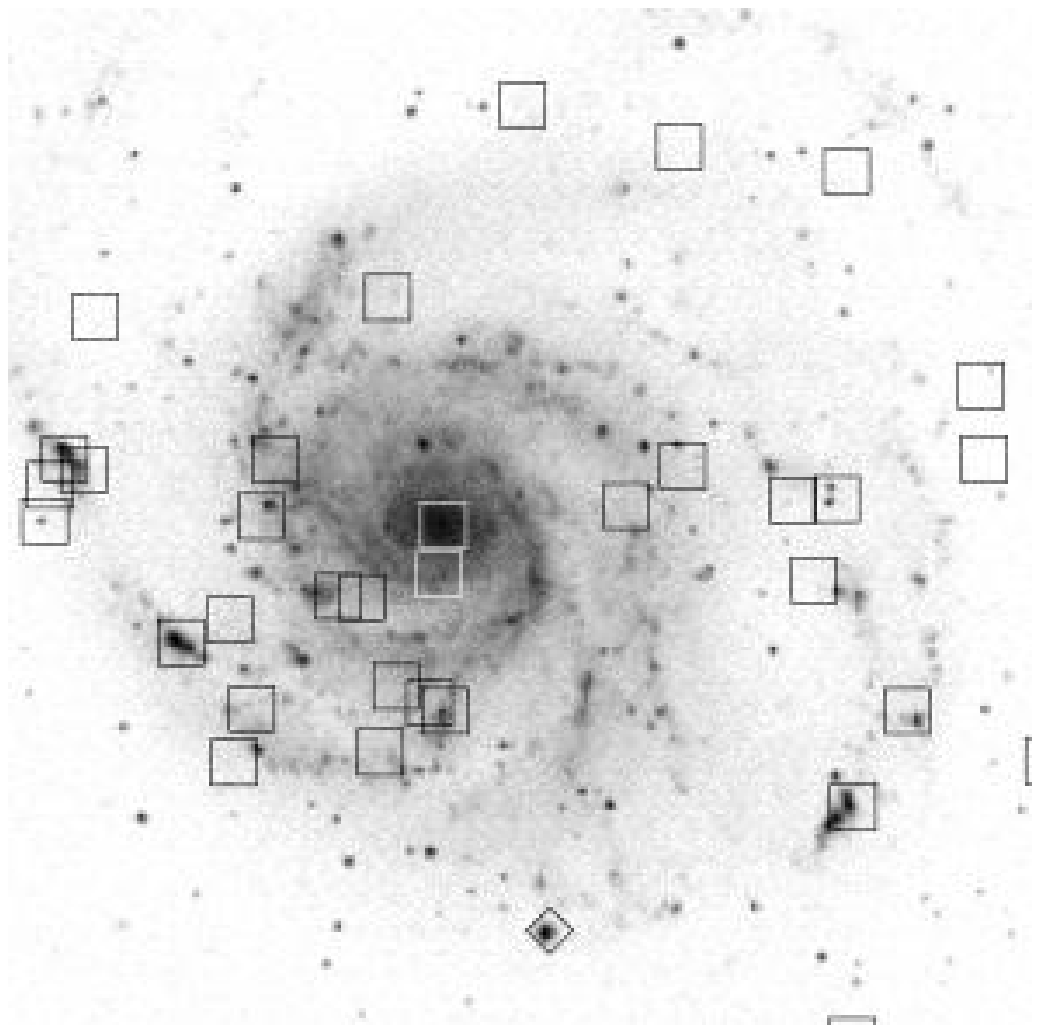}{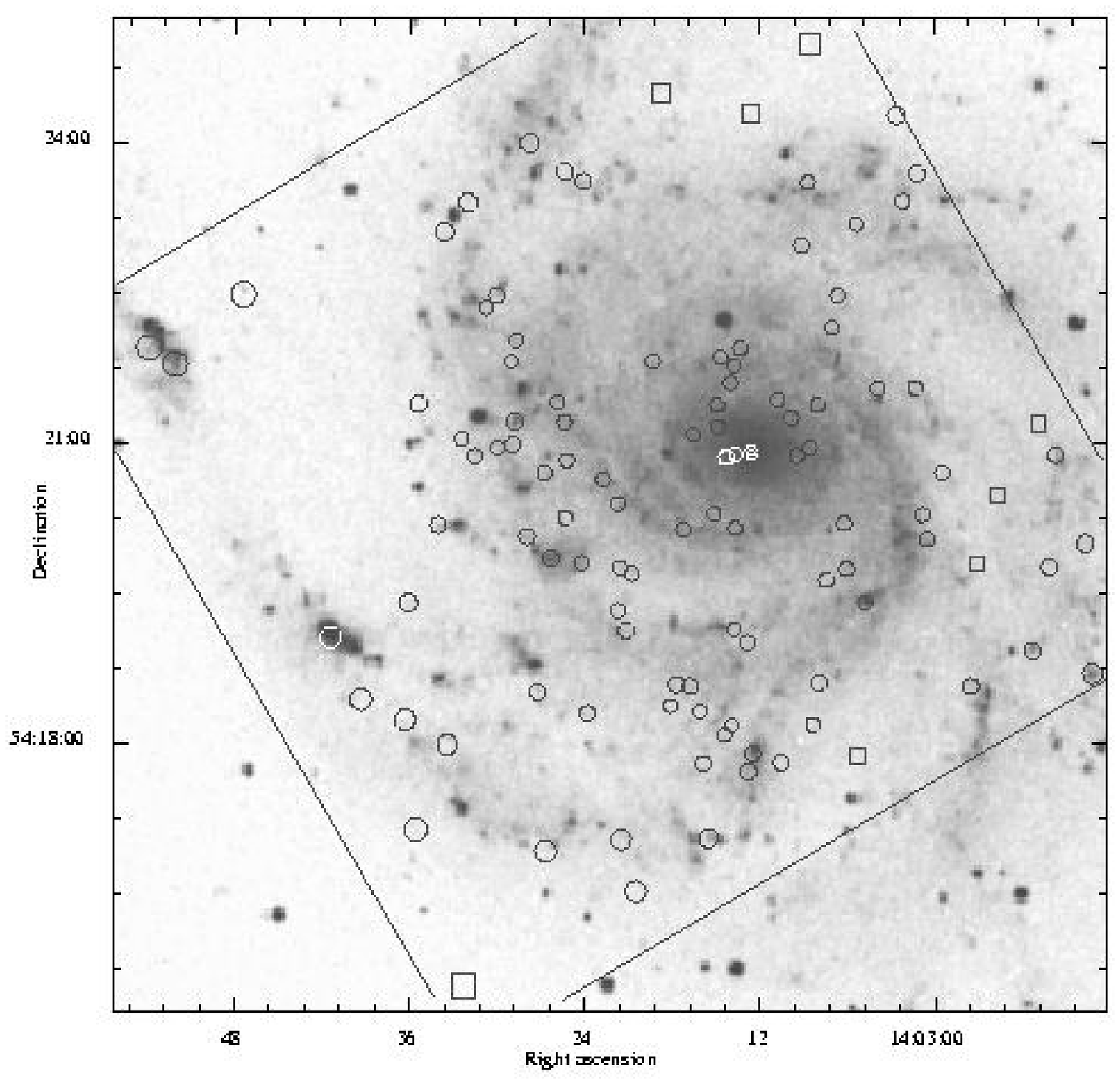}
\caption{Left panel: ROSAT HRI detected sources marked on an optical
image of M101 (Wang et al 1999).  Right Panel: Chandra sources in the
same area (Pence et al 2001.}
\end{figure}

Chandra is an excellent instrument for nearby galaxies.  It's
excellent angular resolution reduces confusion and also results in
lower background leading to better detection sensitivity.  It allows us
to obtain detailed inventories of X-ray sources in galaxies beyond the
Local Group for the first time.  We are able to classify X-ray sources
as binaries, supernova remnants and HII regions by examinging their
X-ray properties and looking for optical/radio counterparts.
Extending high resolution X-ray studies to galaxies beyond the Local
Group allows us to determine how properties of X-ray sources change
with star formation rate and star formation history.   This in turn
will lead to improved theoretical understanding of how X-ray sources
form and evolve.

\section{Bright Globular Cluster X-ray Sources in M31}
Approximately  900 arcmin$^2$ of M31 was surveyed by Chandra in AO2,
mostlt by the HRC GTO team (PI Steve Murray) and Rosanne Di Stefano.  One of the most interesting results to emerge from this census
is that the most luminous X-ray sources in the survey are associated
with globular clusters.  This analysis has been carried out by Rosanne
Di Stefano and her colleagues (see Di Stefano et al 2001).  Of the 30 known globular clusters in the
survey fields, 17 are new.  One third of these globular cluster X-ray
sources have luminosities L$_x>$10$^{37}$ erg s$^{-1}$ .  This is in
contrast to Milky Way globular clusters where only about 10\% have
X-ray luminosities  L$_x>$10$^{37}$ erg s$^{-1}$.  The brightest of
the M31 glaobular cluster sources in the Chandra survey is Bo 375, which has
an X-ray luminosity  L$_x\sim$2-6$\times$10$^{38}$ erg s$^{-1}$, and a
16 hour period (probably orbital).  The luminosity of this source is
a  magnitude larger than the brightest galactic globular
cluster X-ray source, and larger than the Eddington Limit for a 1.4
${\rm M}_{\odot}$ neutron star.

The fact that M31 has more luminous globular cluster X-ray sources
than the Milky Way has long been established (e.g. Primini et al 1993,
and the discussion by Di Stefano et al 2001) and is illustrated
in Figure~3, which shows the cumulative luminosity functions for the
Milky Way and M31.  However, the reason for
the difference has been controversial.  One possibility is simply that M31
has more globular clusters than the Milky Way (the population could be
20 times larger) and hence the probability of finding high luminosity
sources is greater.  It is also possible that there really is
an excess of high luminosity X-ray sources in M31 globular clusters
compared to Milky Way clusters.   Several possibilities are considered
for the origin of the brightest sources
by DiStefano and her collaborators, including that the sources may be
composites, beamed, that M31 clusters have lower metallicty than Milky
Way globular clusters, or that
that they contain accreting black holes.    

One especially interesting suggestion is that the brightest
sources are examples of thermal-timescale mass transfer onto a neutron
star.  This will occur when the donor star expands to fill its Roche
lobe, leading to transfer onto the compact object via the L1 point.
The transfer rate will be very rapid, resulting in  super-Eddington
luminosities. Thermal-timescale mass transfer 
from a high mass donor star onto a black hole (with modest beaming) has been suggested as a
possible mechanism for producing the very high (L$_x>$10$^{39}$ erg
s$^{-1}$) luminosities seen in so-called ``ultra-luminous'' sources in
starburst galaxies (King et al 2001).  DiStefano et al find that they
can explain the 16-hour period of Bo 375 and its high luminosity with
an episode of thermal-timescale mass transfer from a donor star in the
range 1.1-1.6${\rm M}_{\odot}$.  This value is extremely interesting,
and may hold the clue to explaining at least some of the differences
between the X-ray sources in Milky Way and M31 globular clusters.  It
is very unlikely that Milky Way globulars contain many potential donor
stars of this mass, since stars with masses $\ge$0.8${\rm M}_{\odot}$
have evolved off the main sequence.  The globular clusters in M31 may
be younger, in which case they would have many more higher mass donor
stars available.

\begin{figure}
\caption{Cumulative luminosity functions for the globular cluster
X-ray sources in M31 observed with Chandra (solid line) and the Milky
Way (from Verbunt et al 1995)}
\plotfiddle{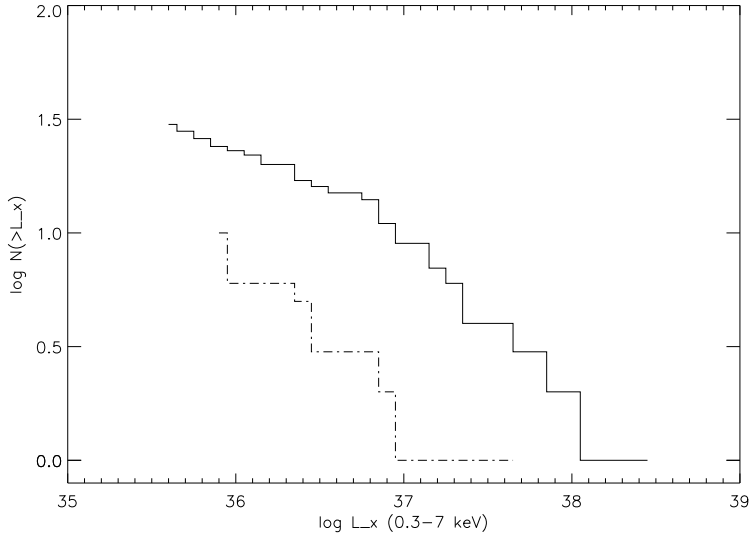}{2.75in}{0.0}{60}{60}{-180.0}{-230.0}
\end{figure}

\section{Hypernova Remnants in M101?}
\label{hyper}

One of the most intriguing suggestions to emerge from ROSAT
observations of M101 is that 5 known optical supernova remnants
have very high X-ray luminosities (L$_x$=$10^{38}$-$10^{39}$ ergs 
s$^{-1}$).  Wang et al (1999) suggested that these highly luminous
might be the remnants of ``hypernovae'' - the energetic event
(e.g. massive star callapsing) responsible for gamma ray bursts.  

Three of these remnants have been observed by Chandra, as reported by
Snowden et al.  One of the
X-ray sources is highly variable and therefore unlikely to be a supernova
remnant.  One of the sources was found (with Chandra's improved
spatial resolution) not to be associated with the optical remnant.  The
last X-ray source was resolved into 2 sources, one of which is
coincident with the optical remnant, but has a ``normal'' X-ray
luminosity.  The second source is a bright nearby binary. 

\section{Observational Properties of X-ray Sources}

\subsection{Optical/X-ray Comparisons}

Chandra observations of nearby spiral galaxies are revealing patterns
similar to that observed by ROSAT for Local Group galaxies.  Sources
can be divided broadly into ``bulge' and ``disk'' components.  Bulge
sources typically cluster around a central nuclear source (which may
be resolved and is often extended) and disk sources trace the spiral
arms.  This pattern is illustrated in Figure~4, which
shows detected X-ray sources plotted on an optical image of NGC 2681
(mostly bulge) and NGC 3184 (mostly disk).  Similar patterns have been
observed for M101 (Pence et al 2001) who find that essentially all
the sources lie along the spiral arms; interarm sources are consistent
with background AGN.  Tennant et al 2001 find that 7/10 of the
brightest sources in M81 fall in the disk spiral arms.  The bulge
sources in M81 follow the I-band profile.   Kaaret (2001) detected 6
sources in M100, all of which lie in the spiral arms.   Blanton et al (2001) show
that the SO galaxy NGC 1553 follows the classic ``bulge'' pattern.

\begin{figure}
\caption{Greyscale optical images of the SO galaxy NGC 2681 (left) and
the the late spiral NGC 3184 (right) with detected X-ray sources.}
\vskip 0.2in
\plottwo{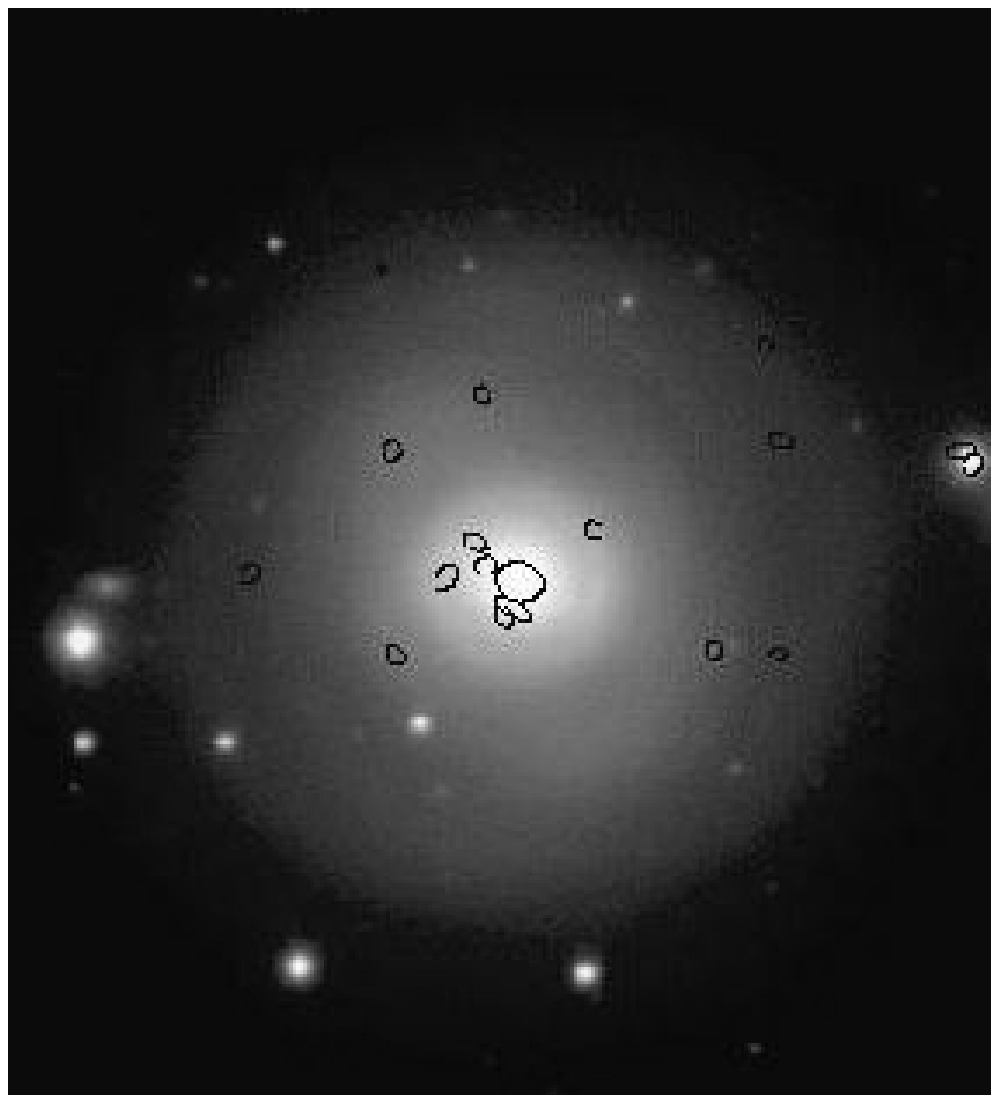}{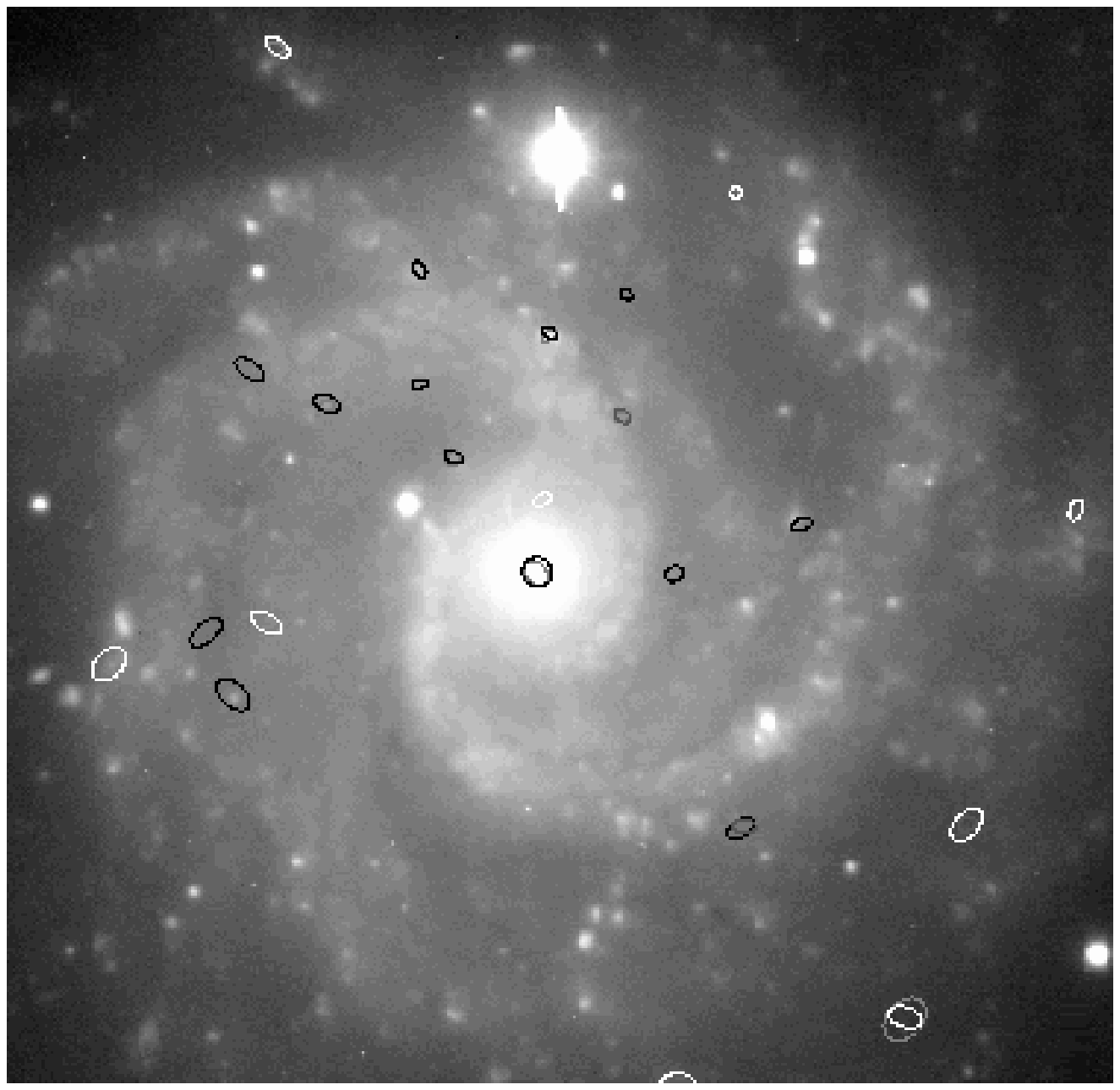}
\end{figure}

\subsection{Variability}

Many Chandra sources in nearby galaxies show evidence for
variability, as might be expected from previous Einsten/ROSAT
observations.  Multiple Chandra observations of NGC 3184 and NGC
2681 reveal that approximately 40\% of the sources 
are variable on scales of 1-4 months.  In M100 4/6 sources detected
by Kaaret (2001)  and one source in Circinus (Bauer et al 2001) show variability
compared to archival data from other missions.  Short term variability
(i.e. within a Chandra observation) has also been observed.  In
Circinus 4 sources exhibit short term variability, 5 in NGC 1553 and
one (very soft) source in M81 (Tennant et al 2001).

\subsection{Spectra}

Figure~5 shows the X-ray color-color diagram for sources in
NGC 1291, NGC 2681, NGC 3184 and NGC 5236.  All of these galaxies are
face-on, thus minimising the effects of extinction.  NGC 1291 and NGC 2681 are
bulge dominated systems whose sources are plotted as triangles. NGC
3184 and NGC 5236 are disk galaxies and are plotted as filled circles.
The Chandra bandpass has been split into three bands: soft (0.3-1.0
keV), medium (1.0-2.0) and hard (2.0-8.0).  The hard color is defined
as $(hard-medium)/total$ and the soft color as $(medium-soft)/total$.  The
spectrum of a source becomes harder as it moves to the right and up
in this diagram.

Also plotted in Figure~5 are the predicted colors of
power-law spectra with increasing photon index (thick solid line).
The tick marks on the thick line mark increasing values (in steps of
0.2) of the photon
index.  The lowest value photon index (hardest)  plotted is 0.7 and the highest
value (softest) is 3.0. Finally, the effect of adding absorption to a
simple power law is illustrated with the thin solid lines (minimum
added absorption is  $N_H=10^{21}$ cm$^{-2}$ and maximum is $N_H=10^{22}$ cm$^{-2}$).   
It is clear from this diagram that most of the sources, whether from
bulge or disk dominated systems have colors that are consistent with
power law spectra plus absorption.  This is futher illustrated in
Figure~6, which shows colors for sources in M82
(diamonds) and NGC 3184 (crosses).  The X-ray sources in M82 have
 colors which are consistent with higher extinction, as might be
expected for an edge-on galaxy compared to a face-on spiral (NGC 3184).
 There is a population of sources
(mostly in the disk galaxies) that have spectra which are very soft
(soft color $<$ -0.5).  It is not clear whether these soft sources are
not found in bulge systems, or whether the detection limit for soft
sources is higher in bulges because of a higher background of diffuse
emission.

These broad conclusions - that the spectra of most sources in spiral
galaxies are consistent with a powerlaw spectrum plus absorption and
that there is a population of soft sources -- are
supported by work by other groups.  Bauer et al and Pence at al (2001)
find that a powerlaw plus absorption is a good description for sources
in Circinus and M101 respectively.  Tennant et al (2001) find that
sources in M81 are described by a $\gamma$=1.6 powerlaw spectrum with
galactic column.  Finally a population of soft sources is observed in
M81 (Tennant et al) and M101 (Pence et al).  Blanton et al do not
observe soft sources in NGC 1553, but the limiting flux is rather high
(10$^{38}$ erg s$^{-1}$).

\begin{figure}
\caption{X-ray color-color diagram for two bulge dominated galaxies
and two disk dominated galaxies.  See text for details.}
\plotfiddle{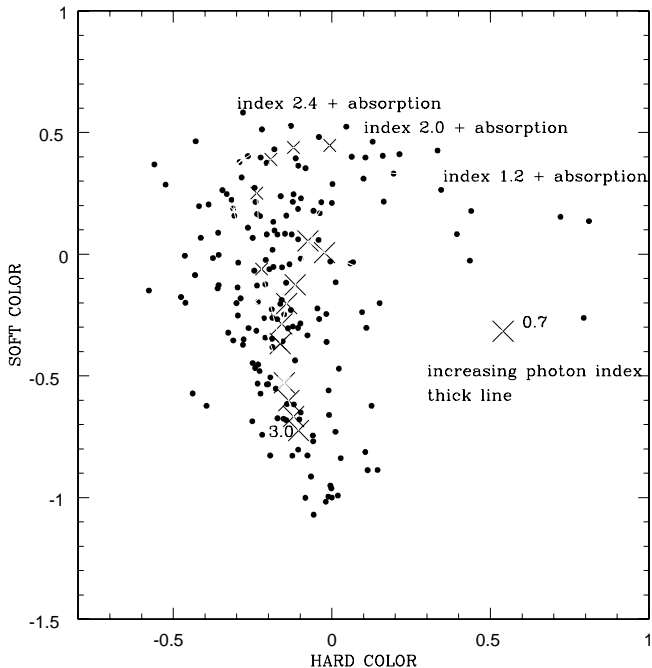}{3.5in}{0.0}{50}{50}{-180.0}{-90.0}
\end{figure}

\begin{figure}
\caption{X-ray color-color diagram for an edge-on starburst galaxy
(M82) and a face-on disk galaxy (N3184).}
\plotfiddle{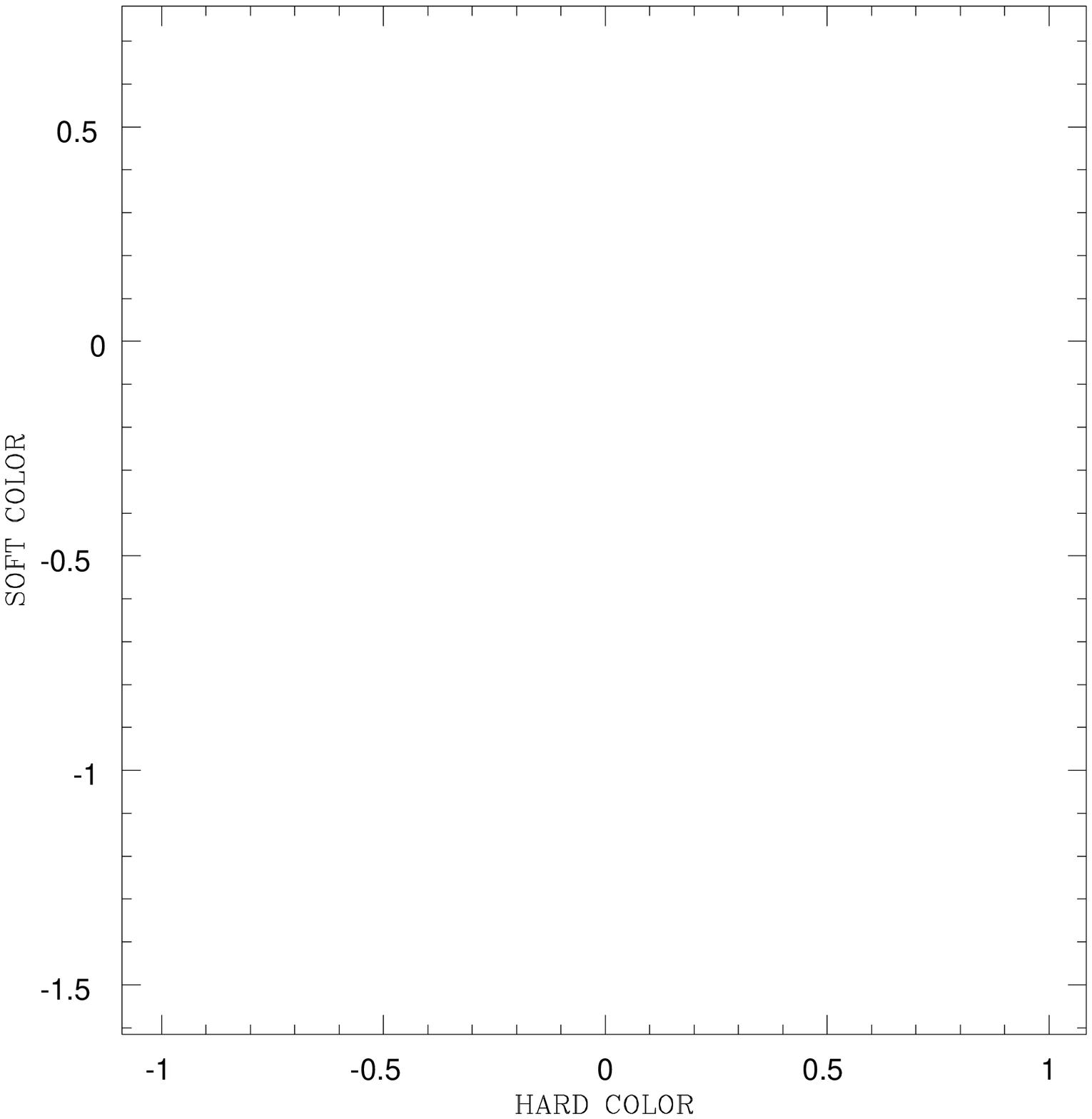}{3.5in}{0.0}{50}{50}{-180.0}{-90.0}
\end{figure}

\subsection{Observational Summary}

The observations described in this section suggest that the discrete
source population in spiral galaxies is dominated by binaries.  This
interpretation is consistent with the colors and variability
characteristics of the sources.  In disk galaxies the sources follow
the spiral arms; these may be short-lived high mass X-ray binaries in
star forming regions.  It is tempting to suggest that the low-luminosity soft sources
found in so many systems may be supernova remnants.  However, at least
some of these sources must be accretion powered because they are
variable.

\section{X-ray Luminosity Functions}
\label{LFs}

The traditional way to compare X-ray source populations in different
galaxies is to look for differences in the cumulatitive luminosity
function. This is the number of sources with a luminosity greater than
a given luminosity ($L$) plotted against $L$. The
luminosity function for two spiral galaxies, NGC 3184 and IC 5332, is
shown in  Figure~7.  The line is the best fit single power-law
to the NGC 3184 data.  Galaxies with luminosity functions with flat
slopes have a relatively large number of high luminosity sources.

\begin{figure}
\caption{Luminosity functions for two spiral galaxies, NGC 3184 and IC
5332}
\plotfiddle{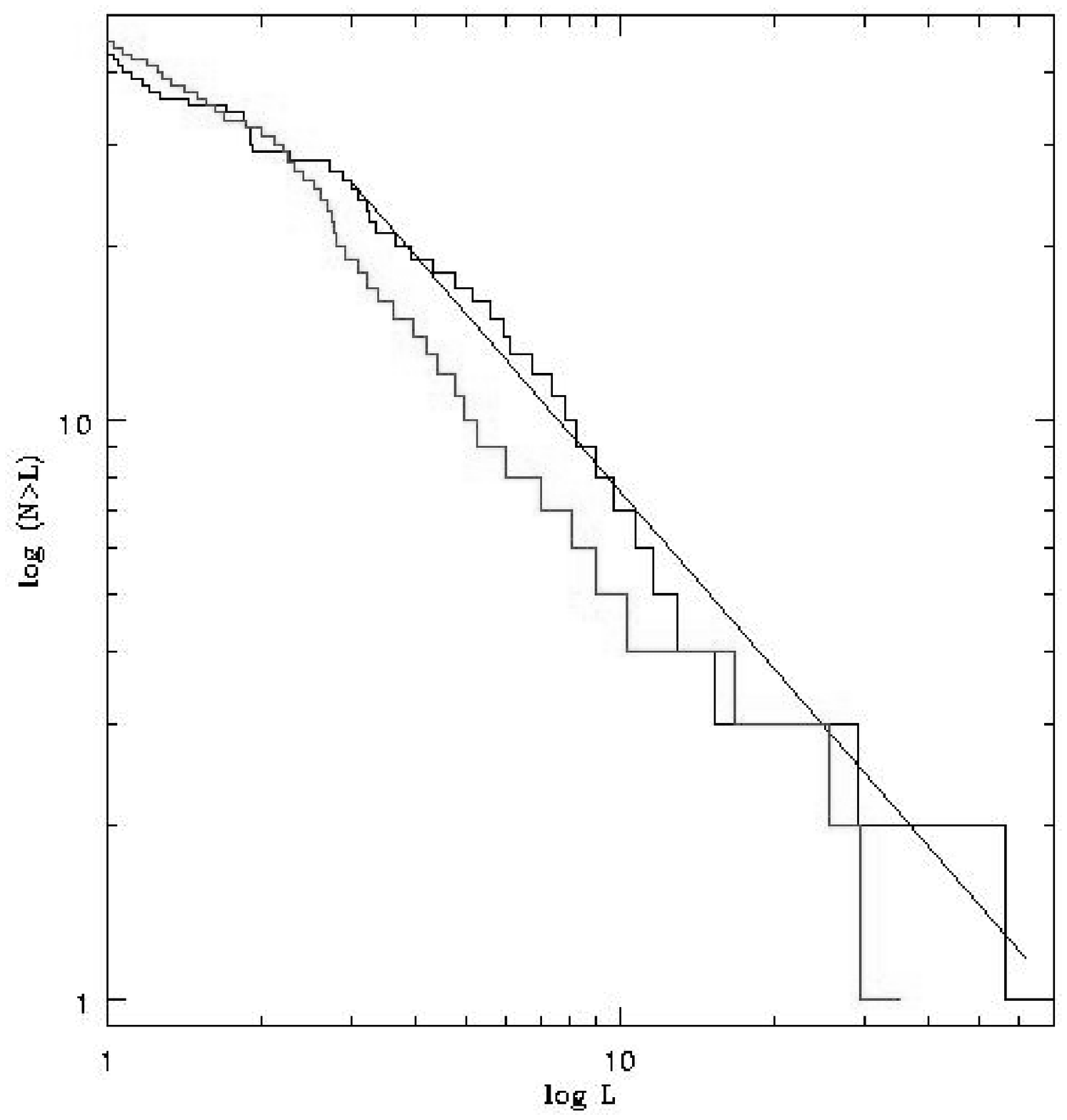}{3.5in}{0.0}{50}{50}{-180.0}{-70.0}
\end{figure}

Slopes have been measured for the luminosity functions for several
galaxies.   These are summarised in Table~1.  Included in
this table are the disks of four classic spirals (M81, M101, IC 5332
and NGC 5236), one starburst (NGC 4038) and three early type/bulge
systems (NGC 1553, the bulge of M31, and NGC 4697).  The M31
luminosity function from Shirey et al uses XMM data.  Two of the
early-type systems have breaks in the luminosity functions (NGC 1553
at 4 $\times 10^{38}$ ergs s$^{-1}$ and NGC 4697 at 3.4 $\times
10^{38}$ ergs s$^{-1}$.)  
\begin{table}
\centering
\begin{tabular}{lll}
\hline
Galaxy &LF slope&ref\\ \hline
\multicolumn{3}{c}{Disks/star forming galaxies}\\  \hline
M81 & 0.5 & Tennant et al\\
M101 & 0.8 & Pence et al\\
IC 5332 & 1.1 & this work\\
NGC 5236 & 0.64 & this work\\
NGC 4038   & 0.45 & Fabbiano et al \\ \hline
\multicolumn{3}{c}{Bulges}\\ \hline
NGC 1553 &1.7& Blanton et al\\
M31 & 1.79 & Shirey et al 2001\\
NGC 4697 & 1.76 & Srazin et al 2000\\  \hline
\end{tabular}
\end{table}

The most striking thing about these results
is that the disks and star forming galaxies seem to have systematically flatter slopes than the
bulges.  A possible caveat is that the detection limit for NGC 1553
is  2$\times10^{38}$ erg s$^{-1}$, so that the luminosity range for
this galaxy does not overlap very well the other galaxies discussed
here.    If this observation is correct, however, it implies that the disks have a larger fraction of higher
luminosity sources relative to the total than do early type systems.
This can be most naturally explained if systems with on-going high
mass star formation have a population of high mass X-ray binaries that
dominate the high end of the luminosity function.   This hypothesis is
supported by the plot in Figure~8, which shows the slopes of
the luminosity functions of several spirals plotted against the
integrated 60$\micron$ luminosity.  There is a clear correlation
between the 60$\micron$ luminosity (a measure of the star formation
rate) and the slope of the X-ray
luminosity function, in the sense that flatter slopes have higher star
formation rates.

\begin{figure}
\caption{The slope of the X-ray luminosity function plotted against
the 60$\micron$ luminosity }
\plotfiddle{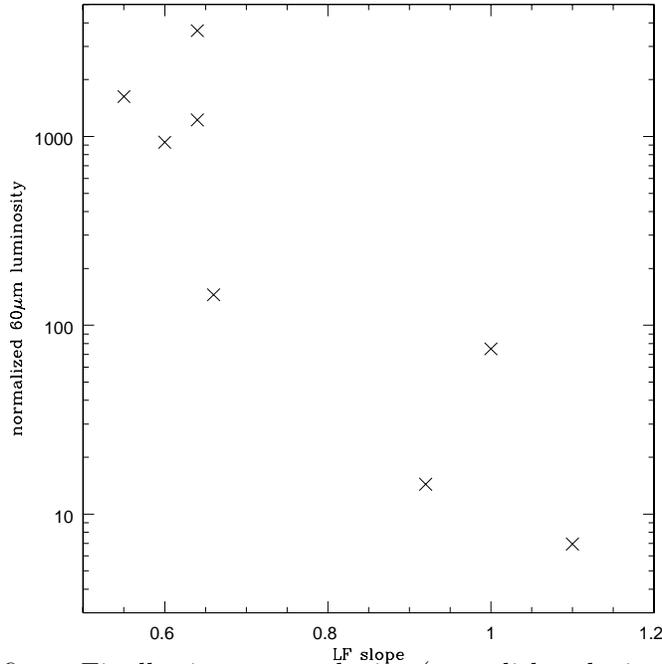}{3.0in}{0.0}{50}{50}{-180.0}{-110.0}
\end{figure}

\section{Simple Population Models}

The luminosity functions described above can be qualitatively
understood in terms of simple binary population models.  As an
example, consider the models of Wu (2001).  He describes binary
formation from three populations: an impulsive starburst event,
constant binary formation, and primordial binary formation.
Primordial binary formation occurs at a low rate throughout the
lifetime of the galaxy as low-mass stars in binary systems  evolve and
their orbits decay, initiating mass transfer.  These systems tend to
be low luminosity (L$_x\sim$10$^{36}$ erg s$^{-1}$) and so do not
contribute significantly to the sources detected by Chandra in nearby
galaxies.  An impulsive star formation event will generate
 many binary systems with about the same age; initially
many of these will be high mass binaries.  This binary population will
evolve with the most luminous systems dying first.  Finally, in many
galaxies (e.g. disk galaxies with star formation occuring in the
spiral arms) binaries will be forming continuously.

Figure~9 shows a simulated luminosity fucnction from Wu
2001.  The overall normalization is set to the luminosity of M81.  The
solid line shows the ``observed'' luminosity function.  The dotted
line shows a binary population formed in a starburst event, the dashed
line shows binaries that are coninually forming.  In the left panel
the most luminous binaries  formed in the starburst episode have
already turned off; the systems that are forming now dominate the high
 end of the luminosity function.  In the right panel the
starburst population is relatively unimportant and the ``constant star
formation'' population dominated the luminosity function.

These models can be extended to other galaxies.  The luminosity
functions in young starbursts like the Antennae (NGC 4038) will
reflect the birth luminosity function, and will show little or no aging.
Continually forming binaries will dominate the luminosity functions in
disk galaxies.  If bulges and elliptical galaxies are formed via
mergers and starbursts then (as suggested by Wu 2001) we might expect to see ``breaks'' in the
luminosity function corresponding to the last major star formation
event.  Breaks in the luminosity functions have been observed in early
type systems, as described in Section~\ref{LFs}.  These breaks may be due to
past starburst episodes, or they may be due to the transition from
neutron star binaries to black hole binaries.  Much more work is
necessary to determine whether the break at  4 $\times 10^{38}$ ergs
s$^{-1}$ is universal, and whether breaks are seen at other
luminosities.  

\begin{figure}
\caption{}
\plottwo{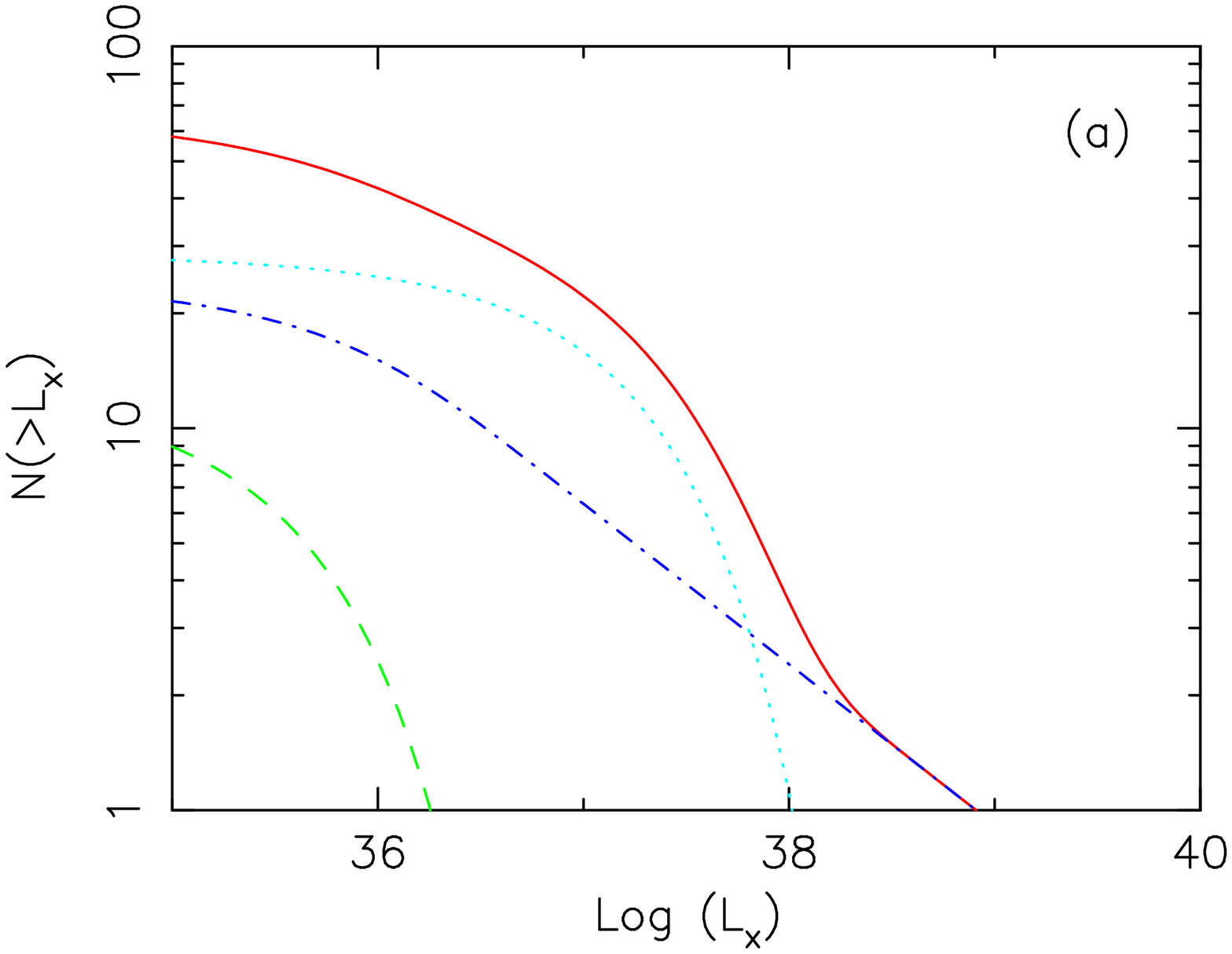}{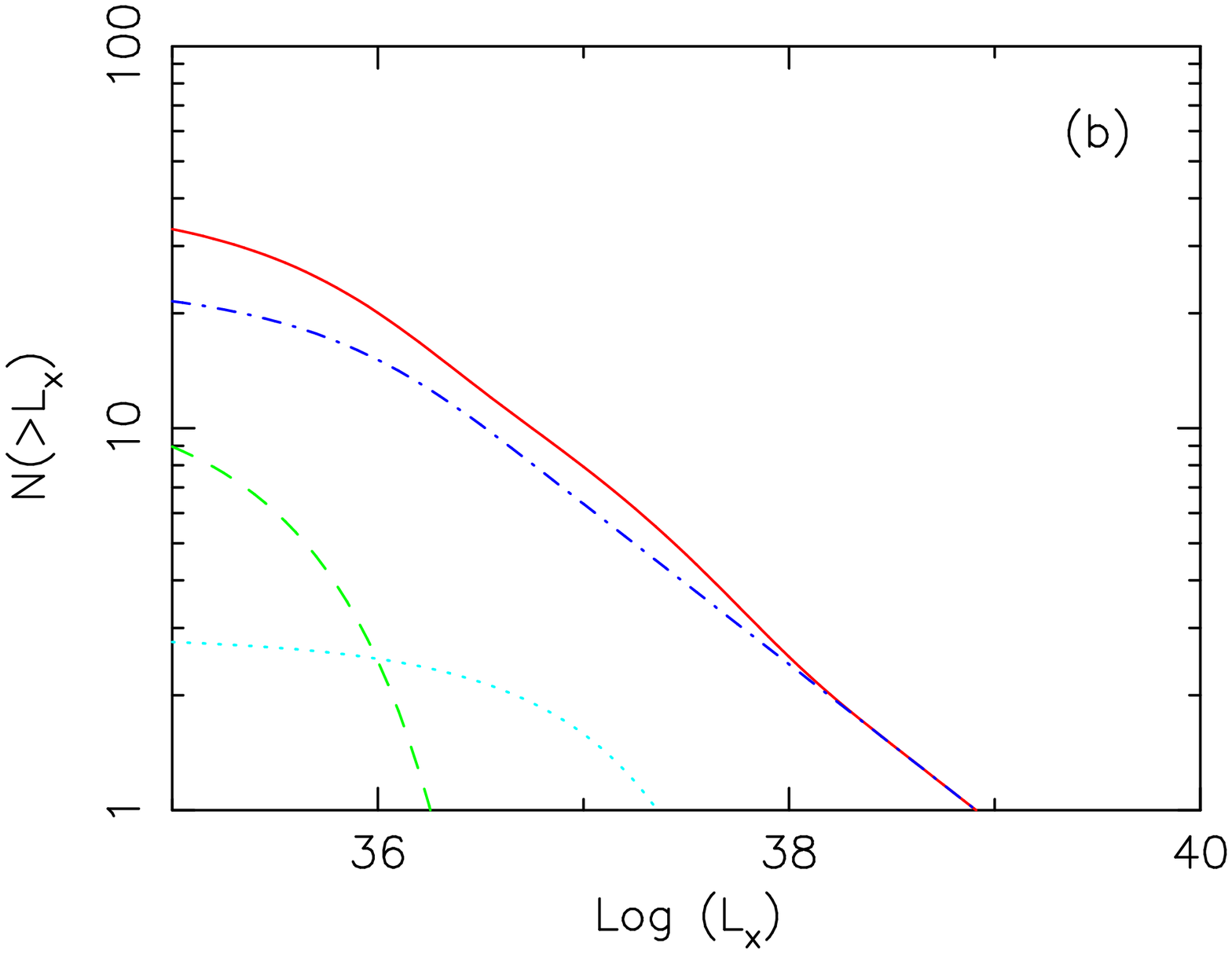}
\end{figure}

\section{Summary and Conclusions}

Chandra's view of nearby spiral galaxies is spectacular, revealing a
multitude of X-ray sources.  In our neighbor M31 Chandra has detected
several (surprisingly) bright X-ray sources associated with globular
clusters, the brighest of which (Bo 375) has
an X-ray luminosity  L$_x\sim$2-6$\times$10$^{38}$ erg s$^{-1}$.  Bo
375 (and possibly other sources)  may be experiencing an episode of
thermal timescale mass transfer, which will allow persistently
super-Eddington luminosities. 

Chandra provides detailed X-ray images of galaxies beyond  the
Local Group for the first time. For example, the  unprecidented positional accuracy of
Chandra allowed Snowden et al (2001) to show that several sources
in M101 cannot be hypernova remnants, as previously suggested from
ROSAT data.  Chandra imaging of several nearby spirals shows that the
X-ray source population is dominated by accreting binaries.  Bright
sources are frequently seen along the spiral arms, suggesting that
these sources are mostly HMXBs.   This conclusion is supported by a
correlation between the slope of the X-ray luminosity function and the
60$\micron$ luminosity.  

 The
luminosity functions of these galaxies can be qualitatively understood
in terms of simple population models.  In disk galaxies binaries are
forming continually, whereas in bulges a ``decaying starburst''
component may be important.  The very flat luminosity functions in
starbursts may refect the birth luminosity distribution.  There is
evidence for breaks in the luminosity functions of early-type
systems.  These may be due to past starburst episodes, or possibly the
transition from neutron stars to black hole binaries. 

\section{Acknowledgments}
It is a pleasure to acknowledge the contributions of many friends and
collaborators, especially Roy Kilgard, Phil Kaaret, Miriam Krauss,
Andreas Zezas, Jonanthan McDowell and Mike Raley.  Thanks also to
Kinwah Wu, Rosanne DiStefano and Doug Swartz for very helpful
discussions.  This work was supported by NASA contract NAS 5839073
(CXC) and Chandra award GO1-2092A.

\end{document}